\documentclass[12pt]{article}
\usepackage{html}
\usepackage{epsf}
\usepackage{graphicx}
\usepackage{amssymb}
\begin{document}
\title{MATTERS OF GRAVITY, The newsletter of the APS Topical Group on 
Gravitation}
\begin{center}
{ \Large {\bf MATTERS OF GRAVITY}}\\ 
\bigskip
\hrule
\medskip
{The newsletter of the Topical Group on Gravitation of the American Physical 
Society}\\
\medskip
{\bf Number 36 \hfill Fall 2010}
\end{center}
\begin{flushleft}
\tableofcontents
\vfill\eject
\section*{\noindent  Editor\hfill}
David Garfinkle\\
\smallskip
Department of Physics
Oakland University
Rochester, MI 48309\\
Phone: (248) 370-3411\\
Internet: 
\htmladdnormallink{\protect {\tt{garfinkl-at-oakland.edu}}}
{mailto:garfinkl@oakland.edu}\\
WWW: \htmladdnormallink
{\protect {\tt{http://www.oakland.edu/?id=10223\&sid=249\#garfinkle}}}
{http://www.oakland.edu/?id=10223&sid=249\#garfinkle}\\

\section*{\noindent  Associate Editor\hfill}
Greg Comer\\
\smallskip
Department of Physics and Center for Fluids at All Scales,\\
St. Louis University,
St. Louis, MO 63103\\
Phone: (314) 977-8432\\
Internet:
\htmladdnormallink{\protect {\tt{comergl-at-slu.edu}}}
{mailto:comergl@slu.edu}\\
WWW: \htmladdnormallink{\protect {\tt{http://www.slu.edu/colleges/AS/physics/profs/comer.html}}}
{http://www.slu.edu//colleges/AS/physics/profs/comer.html}\\
\bigskip
\hfill ISSN: 1527-3431

\bigskip

DISCLAIMER: The opinions expressed in the articles of this newsletter represent
the views of the authors and are not necessarily the views of APS.
The articles in this newsletter are not peer reviewed.

\begin{rawhtml}
<P>
<BR><HR><P>
\end{rawhtml}
%{\bf \Large Contents:}
\end{flushleft}
\pagebreak
\section*{Editorial}

The next newsletter is due February 1st.  This and all subsequent
issues will be available on the web at
\htmladdnormallink 
{\protect {\tt {https://files.oakland.edu/users/garfinkl/web/mog/}}}
{https://files.oakland.edu/users/garfinkl/web/mog/} 
All issues before number {\bf 28} are available at
\htmladdnormallink {\protect {\tt {http://www.phys.lsu.edu/mog}}}
{http://www.phys.lsu.edu/mog}

Any ideas for topics
that should be covered by the newsletter, should be emailed to me, or 
Greg Comer, or
the relevant correspondent.  Any comments/questions/complaints
about the newsletter should be emailed to me.

A hardcopy of the newsletter is distributed free of charge to the
members of the APS Topical Group on Gravitation upon request (the
default distribution form is via the web) to the secretary of the
Topical Group.  It is considered a lack of etiquette to ask me to mail
you hard copies of the newsletter unless you have exhausted all your
resources to get your copy otherwise.

\hfill David Garfinkle 

\bigbreak

\vspace{-0.8cm}
\parskip=0pt
\section*{Correspondents of Matters of Gravity}
\begin{itemize}
\setlength{\itemsep}{-5pt}
\setlength{\parsep}{0pt}
\item John Friedman and Kip Thorne: Relativistic Astrophysics,
\item Bei-Lok Hu: Quantum Cosmology and Related Topics
\item Veronika Hubeny: String Theory
\item Beverly Berger: News from NSF
\item Luis Lehner: Numerical Relativity
\item Jim Isenberg: Mathematical Relativity
\item Lee Smolin: Quantum Gravity
\item Cliff Will: Confrontation of Theory with Experiment
\item Peter Bender: Space Experiments
\item Jens Gundlach: Laboratory Experiments
\item Warren Johnson: Resonant Mass Gravitational Wave Detectors
\item David Shoemaker: LIGO Project
\item Stan Whitcomb: Gravitational Wave detection
\item Peter Saulson and Jorge Pullin: former editors, correspondents at large.
\end{itemize}
\section*{Topical Group in Gravitation (GGR) Authorities}
Chair: Steve Detweiler; Chair-Elect: 
Patrick Brady; Vice-Chair: Manuella Campanelli. 
Secretary-Treasurer: Gabriela Gonzalez; Past Chair:  Stan Whitcomb;
Members-at-large:
Frans Pretorius, Larry Ford,
Scott Hughes, Bernard Whiting,
Laura Cadonati, Luis Lehner.
\parskip=10pt

\vfill
\eject

\section*{\centerline
{we hear that \dots}}
\addtocontents{toc}{\protect\medskip}
\addtocontents{toc}{\bf GGR News:}
\addcontentsline{toc}{subsubsection}{
\it we hear that \dots , by David Garfinkle}
\parskip=3pt
\begin{center}
David Garfinkle, Oakland University
\htmladdnormallink{garfinkl-at-oakland.edu}
{mailto:garfinkl@oakland.edu}
\end{center}

Gary Horowitz has been elected to the National Academy of Sciences

Alessandra Buonanno, Alejandro Corichi, Gabriela Gonzalez, James Hough,  Donald Marolf,
Roger Penrose,  Frans Pretorius, Carlo Rovelli, Madhavan Varadarajan, and David Wands
have been elected Fellows of the International Society for General Relativity and Gravitation.

Hearty Congratulations!

\vfill\eject

\section*{\centerline
{New Links in the International Network}\\ 
\centerline {of Gravitational-wave Detectors}}
\addtocontents{toc}{\protect\medskip}
\addcontentsline{toc}{subsubsection}{
\it Network of Gravitational-wave Detectors, by Stan Whitcomb}
\parskip=3pt
\begin{center}
Stan Whitcomb, Executive Secretary, Gravitational Wave International Committee
\htmladdnormallink{stan-at-ligo.caltech.edu}
{mailto:stan@ligo.caltech.edu}
\end{center}

One of the goals of the Gravitational Wave International Committee (GWIC) is to help foster international collaboration among different parts of the gravitational wave community and to promote the development of gravitational-wave detection as an astronomical tool.  To help identify and support priorities for the international community, GWIC prepared a Roadmap for the field 
(\htmladdnormallink{\protect {\tt{https://gwic.ligo.org/roadmap}}}
{https://gwic.ligo.org/roadmap}).  

For ground-based detectors, the GWIC Roadmap identified as its highest priority:
“The construction, commissioning and operation of the second generation global ground-based network comprised of instruments under construction or planned in the US, Europe, Japan and Australia.”
The motivation is simple: a gravitational wave detector network's ability to locate sources depends directly on the separation of its elements.  The figure below shows the area on the sky into which sources of bursts of gravitational waves can be located with the network  of LIGO and Virgo detectors (L1H1V1H2) compared with networks with a detector in Japan (J1), Australia (A2), or both (A2J1).  The advantage of the global network advocated by GWIC is obvious.

\begin{figure}[h]
   \centering
   \parbox{5.5in}{\includegraphics[clip,angle=0,width=.55\textheight]{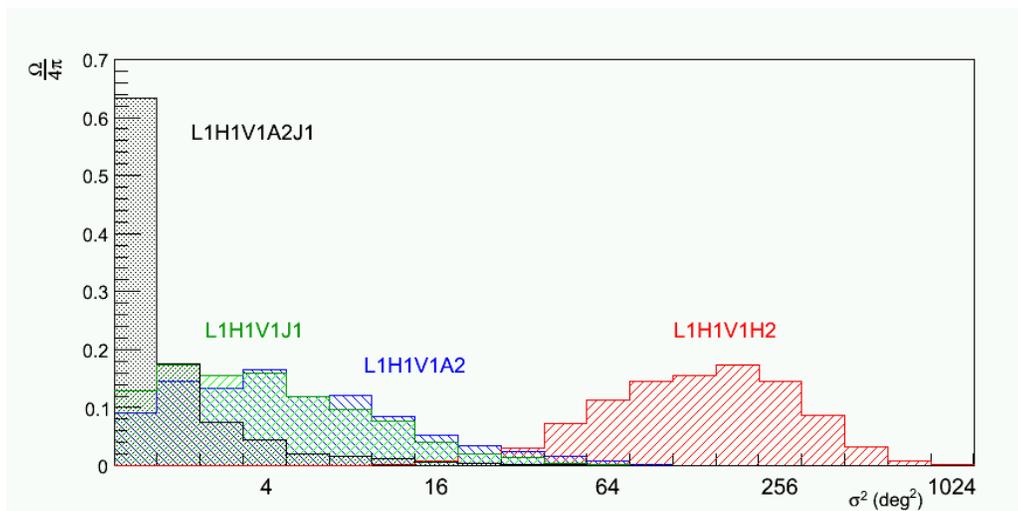}
   \caption{Histogram showing the fraction of the sky producing an angular area error box of a given size for simulated events with SNR ~ 30, observed with different gravitational wave networks.  Graph is taken from Weiss, et al. (https://dcc.ligo.org/cgi-bin/DocDB/ShowDocument?docid=14936).}%
}%
\end{figure} 

In the past year, there have been two exciting developments that significantly improve the prospects to realize this goal.  The currently operating interferometric detectors, LIGO, Virgo and GEO600, have committed to work together to form the core of the international network envisioned in the GWIC Roadmap.  All three projects are engaged in major detector upgrades (Advanced LIGO, Advanced Virgo and GEO HF) designed to propel them into the second generation.  However, these detectors, located solely in North America and Europe, leave much of the desired global network unfinished.  

The first exciting new development comes from Japan.  For the past several years, the Japanese gravitational wave groups have been performing the research and working the design for the Large-scale Cryogenic Gravitational-wave Telescope (LCGT).  In June this year, they learned the good news that LCGT has been funded for phase 1 construction.  LCGT incorporates several innovative features, including an underground location in the Kamioka mine, and cryogenic operation to reduce thermal noise (in phase 2).  When it becomes operational, LCGT will be a crucial link in the international network.

The second new development is LIGO-Australia.  The LIGO Laboratory and the Australian Consortium for Interferometric Gravitational Astronomy (ACIGA) have developed a proposal in which LIGO would deploy one of its Advanced LIGO detectors (currently planned as a second detector for its Hanford Washington site) to a new facility to be built by Australia.  Such a detector would provide the crucial southern hemisphere node needed by the international network.  This proposal passed a major milestone at the end of the summer when the NSF formally accepted the concept, subject to satisfactory management conditions.  A formal proposal to the Australian government for the construction of the facilities is in preparation.  A crucial decision point will come in late 2011, when the LIGO Laboratory will have to decide whether to install the second detector at Hanford, or set it aside for LIGO-Australia.  Unless a commitment to fund the Australian facility can be made by that decision point, LIGO Laboratory will have to install the detector in Hanford.  Despite the obvious challenge of meeting this short deadline, this proposal offers the best opportunity for a second generation interferometric detector in the Southern Hemisphere to date.

\vfill\eject

\section*{\centerline
{New tests of General Relativity}}
\addtocontents{toc}{\protect\medskip}
\addtocontents{toc}{\bf Research briefs:}
\addcontentsline{toc}{subsubsection}{
\it New tests of General Relativity, by Quentin Bailey}
\parskip=3pt
\begin{center}
Quentin Bailey, Embry-Riddle Aeronautical University 
\htmladdnormallink{baileyq-at-erau.edu}
{mailto:baileyq@erau.edu}
\end{center}

The last decade has seen 
a rapid increase in the number
of precision tests of relativity. 
This research has been motivated by the intriguing possibility that 
tiny deviations from relativity might arise in the
underlying theory that is widely believed to successfully mesh
General Relativity (GR) with quantum physics 
\cite{tables,LVreviews}.
Many of these tests have been analyzed within an effective field theory 
framework which generically describes possible 
deviations from exact relativity 
\cite{sme1}
and contains some traditional test frameworks as limiting cases 
\cite{km09}.
One part of the activity has been a resurgence of interest in tests
of relativity in the Minkowski-spacetime context,
where Lorentz symmetry is the key ingredient.
Numerous experimental and observational constraints
have been obtained on many different types of relativity deviations 
involving matter \cite{tables}.
Another part,
which has developed more recently,
has seen the effective field theory framework 
extended to include the curved spacetime regime \cite{akgrav},
and recent theoretical work within this framework has shown 
that there are many unexplored ways in which 
the foundations of GR can be tested 
\cite{qbkgrav,tkgrav}.
Qualitatively new signals for deviations from local Lorentz symmetry 
involving lunar laser ranging observations \cite{llr} 
and atom interferometry experiments \cite{atom} 
have already been analyzed within this framework,
and many exciting new possibilities exist for future work including
proposed Weak Equivalence Principle (WEP) tests.

In the context of effective field theory in curved spacetime, 
relativity violations of these types can be described by
an action that contains 
the usual Einstein-Hilbert term, 
a matter action, 
plus a series of terms describing Lorentz violation for gravity 
and matter.
One useful limiting case of this construction has an action of the form
\begin{equation}
S_g = 
{\frac {1}{16\pi G_N}} \int d^4x \sqrt{-g} ~R 
\ + \ {\frac {1}{16\pi G_N}} \int d^4x \sqrt{-g}~ s_{\mu \nu} R_T^{\mu \nu} 
\ + \ S^\prime.
\label{gravity}
\end{equation}
In this expression the first term is the conventional 
Einstein-Hilbert action for GR, 
while the second term is the leading
``pure-gravity'' Lorentz-violating coupling 
to the traceless Ricci tensor $R_T^{\mu \nu}$, 
which is controlled by $9$ coefficient fields denoted $s_{\mu \nu}$.
The last term $S^\prime$ includes possible dynamical terms for
the coefficient fields $s_{\mu \nu}$. 
Lorentz violation in the classical point-mass limit of 
the matter sector is described by the action
\begin{equation}
S_M=\int d\lambda \left(-m\sqrt{-(g_{\mu \nu}+2c_{\mu \nu})u^\mu u^\nu}
-a_\mu u^\mu\right),
\label{matter}
\end{equation}
where $u^\mu=dx^\mu/d\lambda$ is the worldline tangent and $c_{\mu \nu}$ and $a_\mu$ 
are the coefficient fields that control local Lorentz violation 
for matter.  
In contrast to $s_{\mu \nu}$, 
these coefficients depend on the type of point mass 
(particle species) and so they can also violate WEP.
Perfect local Lorentz symmetry for gravity and matter
is restored when the coefficients $s_{\mu \nu}$, $c_{\mu \nu}$, 
and $a_\mu$ vanish.

It turns out that consistency with Riemann geometry
imposes the requirement that these relativity violations arise 
via so-called spontaneous Lorentz-symmetry breaking
\cite{akgrav}.
In this scenario,
tensor fields in the underlying theory spontaneously
acquire background values through a dynamical process.
In the context of the pure-gravity and matter-gravity 
couplings in \ref{gravity} and \ref{matter}, 
the coefficient fields 
$s_{\mu \nu}$, 
$c_{\mu \nu}$, 
and $a_\mu$ can then be expanded around their background values 
${\bar s}_{\mu \nu}$, 
${\bar c}_{\mu \nu}$, 
and ${\bar a}_\mu$.
The spontaneous breaking of Lorentz symmetry
leaves a modified spacetime metric $g_{\mu \nu}$
and modified point-particle equations of motion.
These results have been obtained in the linearized
gravity limit and the results rely only on the vacuum 
values ${\bar s}_{\mu \nu}$, ${\bar c}_{\mu \nu}$, and ${\bar a}_\mu$.
Calculations of observables in the post-newtonian limit
can then reveal the dominant signals for Lorentz violation
controlled by these coefficients.

Several novel features of the post-newtonian limit 
arise in this effective-field framework.
Some of the effects can be matched to 
the well-established PPN formalism \cite{cmw},
but others lie outside it \cite{qbkgrav,phon}.
Thus the effective-field formalism complements and extends
the large body of analysis that exists within the PPN framework, 
revealing new directions to explore 
via the ${\bar s}_{\mu \nu}$, ${\bar c}_{\mu \nu}$, and ${\bar a}_\mu$ coefficients.
The matter coefficients ${\bar c}_{\mu \nu}$ and ${\bar a}_\mu$, 
which can depend on particle species, 
contribute terms directly to the post-newtonian metric.
This implies, 
for example, 
that two (chargeless) point-like sources with the same total mass
but different composition 
yield gravitational fields of different strength.
The coefficients for Lorentz violation ${\bar c}_{\mu \nu}$ and ${\bar a}_\mu$ 
also modify the equations of motion for matter, 
thus implying that they also control WEP violations.

From the post-newtonian metric and the standard geodesic equation 
for test bodies, 
the primary effects due to the nine coefficients ${\bar s}_{\mu \nu}$ 
can be obtained. 
Tests that can measure these coefficients 
include Earth-laboratory tests with gravimeters, 
torsion pendula, 
and short-range gravity experiments.
Space-based tests include lunar and satellite laser ranging,
studies of the secular precession of orbital elements
in the solar system and with binary pulsars, 
and orbiting gyroscope experiments. 
In addition, 
classic effects such as the time delay and bending of light
near a massive body are affected.

Constraints on the ${\bar s}_{\mu \nu}$ coefficients have already been reported.
The main observable effects of Lorentz violation in the Earth-Moon
orbit are oscillations in the lunar range.
For example, 
one such unconventional oscillation occurs at a frequency
of twice the mean orbital frequency, 
and can be traced to the violation of the conservation of 
angular momentum for the two-body system. 
Using lunar laser ranging data spanning over three decades, 
Battat, Chandler, and Stubbs placed constraints on $6$ combinations of the 
${\bar s}_{\mu \nu}$ coefficients at levels of $10^{-7}$ to $10^{-10}$ \cite{llr}.
The local acceleration on the Earth's surface 
becomes modified in the presence of the ${\bar s}_{\mu \nu}$ accelerations, 
leading to $7$ measurable coefficients in 
Earth-laboratory experiments.
These coefficients were measured by M\"uller {\it et al.} using an 
atom interferometer as a (vertical) gravimeter, 
resulting in $7$ constraints at the level 
of $10^{-6}$ to $10^{-9}$ \cite{atom}.
The novel effects of the coefficients ${\bar s}_{\mu \nu}$ would also
result in a horizontal acceleration modulated by the Earth's
sidereal rotation frequency and its orbital frequency.
A suitable interferometer can potentially measure this
effect and may be used for future work.

For bodies interacting gravitationally,
modifications to the standard geodesic
equations of motion result from the coefficients 
${\bar c}_{\mu \nu}$ and ${\bar a}_\mu$, 
which violate Lorentz symmetry and WEP \cite{tkgrav}.
Ground-based gravimeter, 
atom interferometry, 
and WEP experiments are among the existing and proposed tests 
that can probe these coefficients. 
Lunar and satellite laser ranging observations as
well as measurements of the perihelion precession of the planets
are also potentially of interest.

Space-based WEP tests are among the most sensitive tests 
for probing the ${\bar c}_{\mu \nu}$ and ${\bar a}_\mu$ coefficients.
The relative acceleration of two test bodies of different composition
is the observable of interest for these tests.
Some novel time-dependent effects arise
when the relative acceleration is calculated in the satellite 
reference frame in the presence of the coefficients 
${\bar c}_{\mu \nu}$ and ${\bar a}_\mu$. 
In the effective field theory approach to describing 
Lorentz violation, 
the standard reference frame for reporting measurements is 
the Sun-centered celestial equatorial reference frame
or SCF for short.
Oscillations in the relative acceleration occur at a number of 
different frequencies including multiples and combinations
of the satellite's orbital and rotational frequencies, 
as well as the Earth's orbital frequency, 
upon relating the satellite frame coefficients to the SCF.
The extraction of Lorentz-violating amplitudes independent of 
the standard tidal effects can be acheived
with this time dependence, 
and up to $9$ independent combinations of the 
coefficients ${\bar c}_{\mu \nu}$ and ${\bar a}_\mu$ can be measured, 
which goes beyond the commonly used $\delta a/a$ parameterization
for WEP tests.
Future tests of particular interest are the STEP \cite{step},
MicroSCOPE \cite{micro}, and Galileo Galilei \cite{gg} experiments.
These tests offer sensitivities
ranging from $10^{-7}$ GeV to $10^{-16}$ GeV for ${\bar a}_\mu$ coefficients
and $10^{-9}$ to $10^{-16}$ for ${\bar c}_{\mu \nu}$ coefficients.

In conclusion, 
recent theoretical work within an effective field theory 
framework has revealed many new possibilities for testing GR.
This framework systematically categorizes different
types of local Lorentz violations for gravity and matter, 
including some types that also violate WEP.
Calculations of observables in this framework have been performed
and they reveal measurable signals for deviations
from perfect local Lorentz symmetry controlled primarily by 
the coefficients ${\bar s}_{\mu \nu}$, ${\bar c}_{\mu \nu}$, and ${\bar a}_\mu$.
For ordinary matter consisting of protons, neutrons, and electrons, 
there are $39$ independent quantities in ${\bar c}_{\mu \nu}$ and ${\bar a}_\mu$, 
while ${\bar s}_{\mu \nu}$ contains $9$ quantities.
Already, 
$8$ of the $9$ coefficients in ${\bar s}_{\mu \nu}$ have been measured
and the results are so far consistent with GR.
However, 
future tests probing the coefficients ${\bar s}_{\mu \nu}$, 
${\bar c}_{\mu \nu}$, 
${\bar a}_\mu$ may reveal minuscule deviations from GR that would 
provide a signal from an underlying unified theory of physics.

\vfill\eject

\section*{\centerline
{Theory Meets Data Analysis}
\centerline { at Comparable and Extreme Mass Ratios}}
\addtocontents{toc}{\protect\medskip}
\addtocontents{toc}{\bf Conference reports:}
\addcontentsline{toc}{subsubsection}{
\it Theory Meets Data Analysis, 
by Steve Detweiler}
\parskip=3pt
\begin{center}
Steve Detweiler, University of Florida 
\htmladdnormallink{det-at-phys.ufl.edu}
{mailto:det@phys.ufl.edu}
\end{center}

The Perimeter Institute (PI) in Waterloo, with local organizers Eric Poisson and Luis Lehner,
hosted the ``Theory Meets Data Analysis at Comparable and Extreme Mass Ratios'' conference in June, 2010. The combined conference consisted of consecutive annual Capra (gravitational radiation reaction) and NRDA (numerical relativity and data analysis) meetings that ran for  seven days with nearly sixty talks. The consecutive format allowed for easy interactions between these two groups and many lively discussions took place in the PI Black Hole Bistro over coffee in the morning and other refreshments later in the day.

During the course of the meetings we had some distractions. Abraham Harte was describing ``Spin-induced bobbing effects in relativistic systems'' when a slide being projected started ``bobbing'' on the screen. Later we were surprised to learn that this was caused by an earthquake centered north of Ottawa and rated in the ``mid fours.'' Other excitement was regularly caused by World Cup action, where the nationality of the celebrators let the rest of us know how a match was going.

With so many talks spread over one week in Waterloo, it seems silly to just fill the page with names and titles, particularly when the schedule with links to many of the abstracts is located at:\\
\htmladdnormallink{\protect {\tt{http://www.perimeterinstitute.ca/Events/Theory\_Meets\_Data\_Analysis/Schedule/}}}
{http://www.perimeterinstitute.ca/Events/Theory_Meets_Data_Analysis/Schedule/}

You can peruse this list yourself and even see and hear any of the talks. PI has an excellent facility for recording the talks with live action from the speaker accompanied by slides on the side. Viewing a talk on the web might be better than being there: you can either replay the confusing part or fast-forward through it! To enjoy a talk just click on the appropriate title in the ``Schedule'' (link above).

What follows is a pr\'ecis % précis
of the highlights from a limited number of the \textit{many} interesting new ideas and results that were presented during the meeting.
Please accept my apology that my Capra interests have influenced the choice of specific talks described below.

As a member of the Capra community, I am pleased to report that we are reaching the end of a long, difficult adolescence. In the self-force portion of the meeting, a few serious meaningful applications of the gravitational self-force were described that allow for detailed comparisons among each other as well as with corresponding post-Newtonian analyses. The gravitational self-force has arrived.

Nori Sago described ``Gravitational self-force effect on the periapsis advance in Schwarzschild spacetime.'' This work with Leor Barack provides the first truly interesting results from the self-force community, which includes the fractional effect of the self-force on the orbital frequency of the ISCO for the Schwarzschild metric, 
$\Delta\Omega_{\rm {isco}}/\Omega_{\rm {isco}}
= 0.4870(\pm 0.0006)\mu/M$.
Later, Leor Barack gave a through description of ``Computational aspects of the gravitational self-force'' as practiced by the very active and broad Southampton group. This particular talk is not included in the PI archive for some unknown reason. However some of the Barack and Sago work, with the collaboration of Thibault Damour, is shown in Fig.~2. In this plot, $x$ is $(M\Omega)^{2/3}$ which is essentially $M/R$ (the Newtonian limit is on the left, and the ISCO is at $x=.166$) while $\rho$ is the $O(\mu/M)$ part of $(\Omega_r/\Omega_\phi)^2$.
This reveals solid agreement  with post-Newtonian analysis. See eprint 
\htmladdnormallink{grqc/1008.0935}{http://arxiv.org/abs/1008.0935} 
for details.
\begin{figure}%
   \centering
   \parbox{3.5in}{\includegraphics[clip,angle=0,width=.35\textheight]{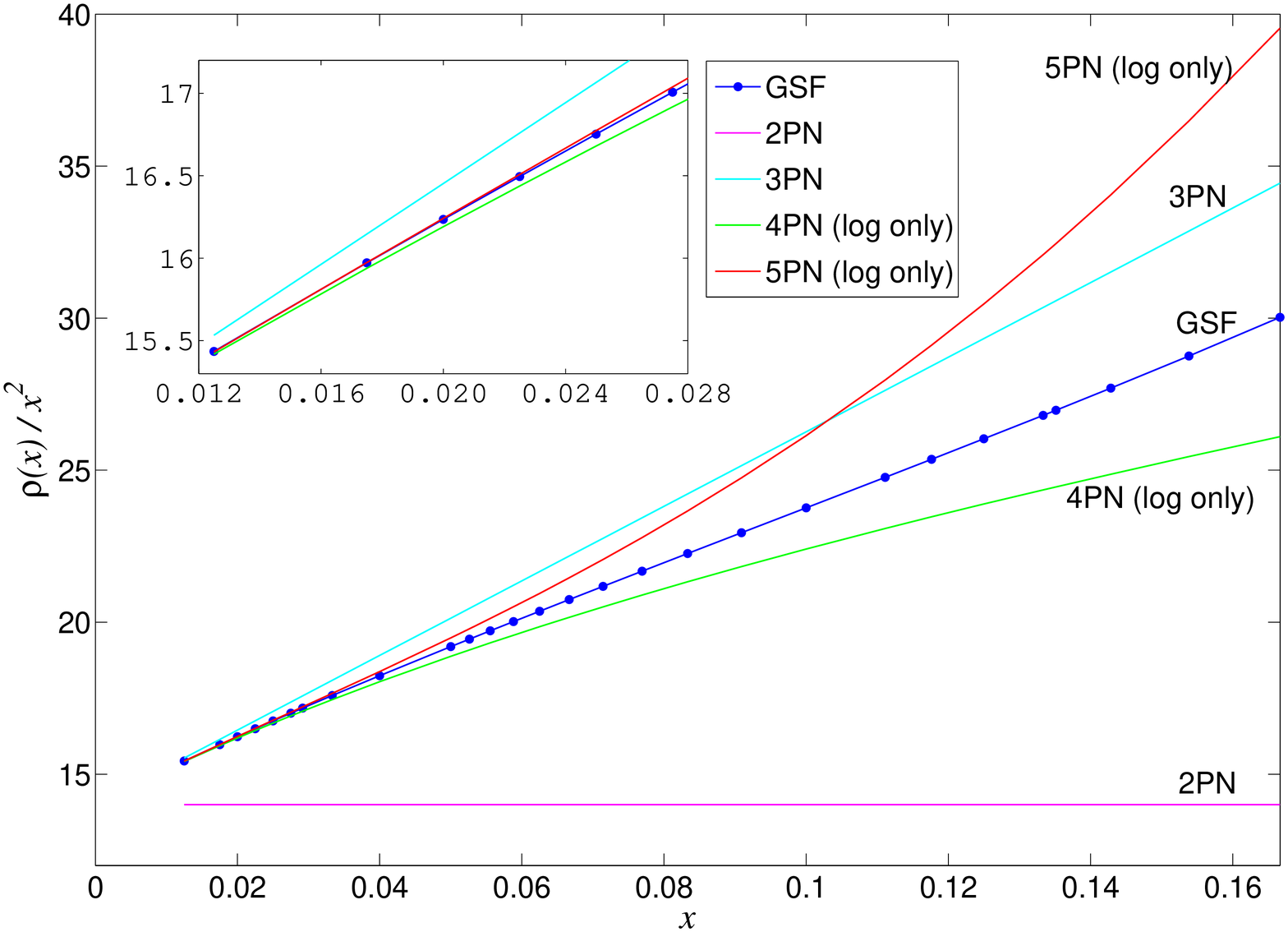}%
   \caption{Periapsis advance, grqc/1008.0935}%
}%
\begin{minipage}{3.5in}%
   \includegraphics[clip,angle=-90,width=.3\textheight]{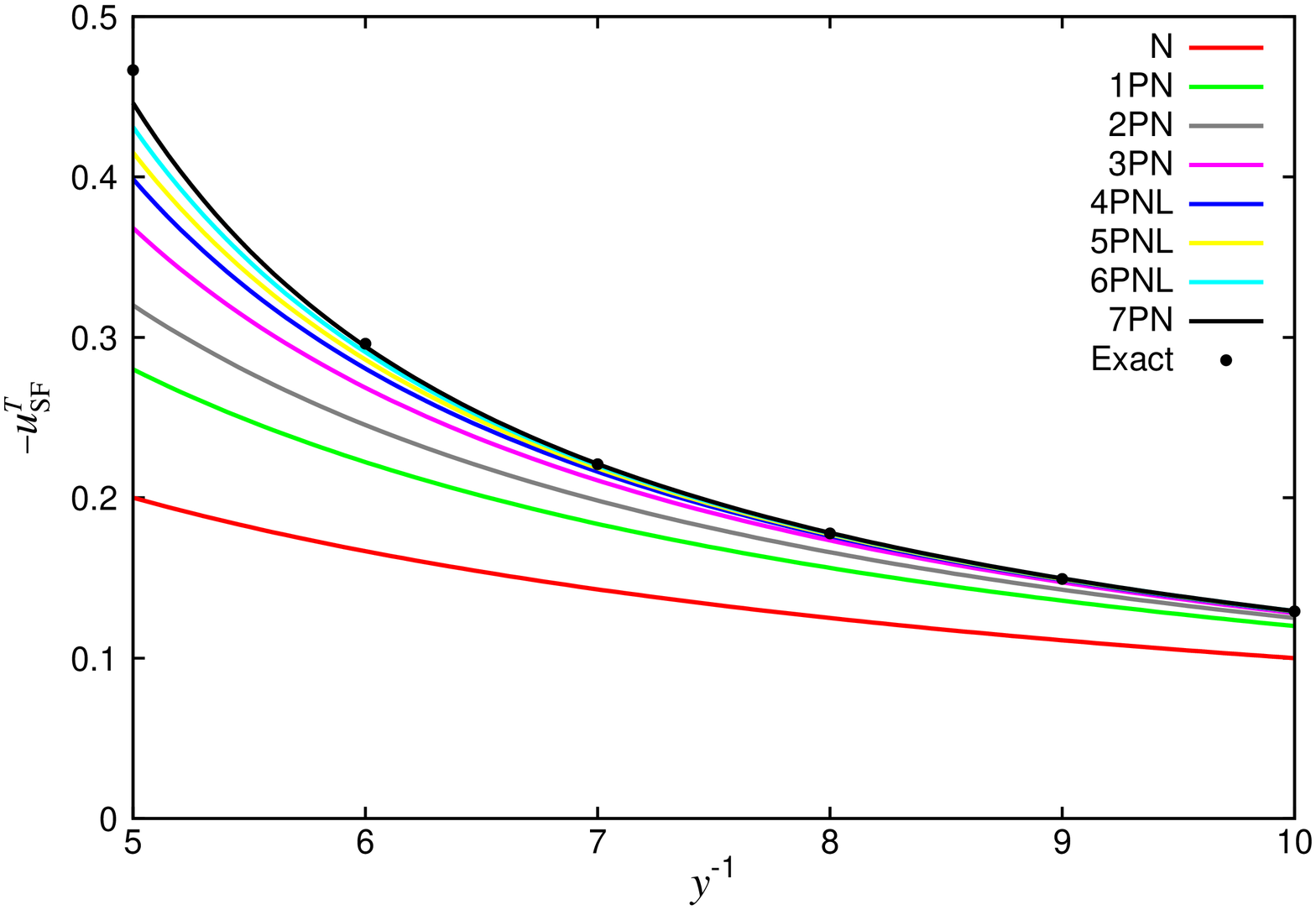}%
   \caption{Redshift observable}%
\end{minipage}%
\end{figure}

Similarly, Alexandre Le Tiec described the general gravitational self-force ``Comparison  with post-Newtonian theory'' and, with Bernard Whiting ``Merging numerical results with post-Newtonian analysis'', covered in some detail the interface between high accuracy numerical self-force results and the high order post-Newtonian analysis for a particular, physically measurable ``redshift'' quantity for circular orbits. The results from a collaboration of Le Tiec and Blanchet with the University of Florida group is presented in Fig.~3, where $y$ is $M/R$, so that the ISCO is at $y^{-1}=6$ and the Newtonian limit is on the right. John Friedman described a Teukolsky-based approach to the same problem. Abhay Shah was responsible for much of the implementation, and the Milwaukee group has obtained results which agree with others to many significant digits.

Ian Vega and Barry Wardell described slightly different approaches to providing non-singular ``effective sources'' to be used in self-force calculations.

Sam Dolan ``Self-force via $m$-mode regularization and time domain evolution,''
Niels Warburton ``Scalar field self-force for eccentric orbits in Kerr,''
and Sarp Akcay ``A fast frequency-domain calculator for the gravitational self force'' described recent developments from the Southampton group. 

In ``Multiscale analysis of extreme mass ratio inspirals in Kerr: separatrix crossing,''
Tanja Hinderer described how an EMRI system could evolve through a transient resonance, where $\omega_r/\omega_\theta=n/m$ for small integers $n$ and $m$. This would result in a dramatic jump in phase error and could be a very significant issue in tracking EMRI phase over many cycles.

Abe Harte presented the ``Foundational aspects of the self-force'' which are now well in hand, at least to the extent that actual calculations of meaningful quantities are now being performed.

Ian Vega described a general technique for studying self-force problems with an extreme mass-ratio which could be implemented via the traditional 3+1 methods of numerical relativity.

Scott Hughes gave an entertaining colloquium style talk
``Probing the physical and astrophysical nature of black holes with gravitational waves'' which left us with the pleasant (but perhaps erroneous) thought that we each understand black holes and gravitational waves very well.

Steve Drasco presented animations of EMRI waveforms based upon a spectral representation known to provide good approximations of
Teukolsky-waveforms. The stroboscopic dancing of the waveforms and the panache of the final chirp should not be missed!
The waveforms come with sound included at
\htmladdnormallink{\protect {\tt{http://www.tapir.caltech.edu/\~{ }sdrasco/animations}}}
{http://www.tapir.caltech.edu/~sdrasco/animations}   

Hiroyuki Nakano described ``Perturbative effects of spinning black holes with applications to full numerical relativity'' which uses a second order perturbative formalism that includes spin effects at the perturbative level with a goal of computing recoil velocities of merging black holes.

Mike Boyle, ``NINJA-2: The Revenge,'' described the current status of the Ninja-2 collaboration which injects numerical relativity waveforms into the LIGO-VIRGO data stream, which is then searched for by the the data analysis community.

Yosef Zlochower described recent results from the RIT group of highly-spinning binaries giving near maximal spins and high-mass ratio binaries. 

Sean McWilliams described the current state of the art of black hole binaries in vacuum but now also including matter, hydrodynamics and E\&M fields.

Parameswaran Ajith summarized the status of modeling the waveforms from binary black holes using calibration from numerical relativity simulations.

Enrico Barausse presented a new EOB model of spinning black hole binaries where the spin effects are are consistent with the test particle limit and also the post-Newtonian limit.

Jocelyn Read, ``Modeling waveforms from binary neutron star,'' described details of waveforms from binary neutron stars which include late-time  modifications to the post-Newtonian waveform models from tidal deformations. In some circumstances a hypermassive remnant produces significant additional signal after the merger.

Zachariah Etienne described the Illinois effort to understand black hole - neutron star mergers while including hydrodynamics and Einstein's equations with a focus on gravitational waveforms and the remnant disk.

Carlos Palenzuela described ``Binary black hole collision in a force-free environment'' which includes electromagnetic fields and plasma and results in the emission of jets.

Sam Waldman, ``Gravitational Wave Detection:  Past, Present and Future,'' sweepingly covered all possibilities for the direct detection of gravitational waves going from very low frequency to high frequency. At the low frequency limit this starts with observations of cosmic microwave polarization and then pulsar timing measurements. Mid-frequency approaches include LISA and other possible space-based efforts. He continued with a focus on initial LIGO and VIRGO and the improvements necessary to increase their sensitivity by a factor of ten in their advanced versions.

In the finale Avery Brodderick, ``Hairstyles of Compact Objects,'' reminded us all that the  real world is messy and that the devil is in the details. The historical failure of detailed models of astrophysical energy sources have sometimes led us to new physics and sometimes to better detailed models. The observations of relatively clean gravitational waves will help clear the air, but we must still confront the mess.

Your correspondent thoroughly enjoyed this combined meeting and looks forward to more of these stimulating events, and he apologizes for either missing or misrepresenting your contribution. At the end of a wonderful week, it was a wistful drive from Waterloo to Niagara and Buffalo, leaving the cool pleasantly rainy weather of Ontario for the oppressive heat south of the border.

\vfill\eject

\section*{\centerline
{Ascona Conference}}
\addtocontents{toc}{\protect\medskip}
\addcontentsline{toc}{subsubsection}{
\it Ascona Conference, 
by Simon Ross}
\parskip=3pt
\begin{center}
Simon Ross, Durham University 
\htmladdnormallink{S.F.Ross-at-durham.ac.uk}
{mailto:S.F.Ross@durham.ac.uk}
\end{center}

A conference on ``Strings, M-theory and Quantum gravity'' was held in
Ascona, Switzerland on the 25th-30th of July 2010. The conference was
organised by Matthias Gaberdiel, Matthias Blau, and Marco Marino in
the beautiful setting of the Monte Verita conference center, on a hill
above the town of Ascona overlooking Lake Maggiore. The conference
center has an interesting history, having served as a site for various
experiments in alternative living in the early twentieth century,
later developing into an important international art and cultural
venue attracting many noted artists and thinkers. It now hosts a range
of academic conferences organised by ETH Zurich, providing well-run
conference facilities and excellent meals in a beautiful setting.

There were a mix of hour-long and half-hour talks covering a range of
topics, with a focus on aspects of AdS/CFT and its extensions. The
organisers did an excellent job of bringing together a good mix of
people to create a stimulating environment. The majority of the
speakers included reports on work in progress, which added to the
interest in their talks and sparked lively discussions. The
discussions were so active that the organisers sometimes had
difficulty getting us to stop in time for lunch!

Slides for the talks are available on the conference website at
\htmladdnormallink{\protect {\tt{http://www.conferences.itp.phys.ethz.ch/doku.php?id=verita10:start}}}
{http://www.conferences.itp.phys.ethz.ch/doku.php?id=verita10:start}

The conference opened with a talk by Hermann Nicolai on arithmetic
quantum gravity, which blended mathematical information on the
octavians (the ring of integer octonions) and studies of the BKL
phenonomenon in cosmology to develop a proposal for quantum gravity
based on $E_{10}$. Jonas Bjornsson next discussed an approach to
supergravity amplitudes based on pure spinors, which predicts a
divergence at seven loops in $d=4$. Later in the day, Martin Cederwall
discussed the use of pure spinors to construct actions for
supergravity with manifest supersymmetry. Masaki Shigemori described
exotic branes with U-duality monodromy, emphasizing that such branes
can appear as supertubes. Later in the week, we heard about the
doubled field theory description of string theory on a torus from
Chris Hull: this is a simplified version of string field theory,
respecting U-duality. Closing the first day, Joerg Teschner explained
the relation between $\mathcal N=2$ supersymmetric field theories and
Liouville theory, and showed that the S-duality invariance of the
field theory is related to known features of the Liouville theory.

There were several talks on aspects of AdS/CFT. On the first day, Ben
Craps reviewed progress on the holographic description of
time-dependent spacetimes. He reviewed work on the description of
singular big-bang like geometries in matrix models, and explained a
new approach which uses adiabiticity to control the late time
behaviour. On the second day, Don Marolf argued that some aspects of
holography are a result of the structure of classical gravity, which
implies that the information available on the boundary at one moment
in time is available at any other moment in time and that in
perturbation theory about anti-de Sitter space, the information on the
boundary at all times is likewise available at any one moment in
time. Veronika Hubeny discussed the bulk description of a quark in
uniform circular motion: this seemed to provide a counterexample to
the usual understanding of the UV/IR relation between the bulk and
boundary, but she showed how this could be understood by thinking of
the radiation in the bulk as a superposition of shock waves. Shiraz
Minwalla discussed perturbative investigations of black holes with
scalar hair in AdS in global coordinates, showing how these black
holes fill a gap in the phase diagram for the theory. Michael Gutperle
described work on the holographic description of CFTs with boundaries,
extending previous work on a pair of theories joined across a boundary
to theories with junctions where multiple CFTs meet.

There were two talks focusing more on the field theory side of the
correspondence. Mukund Rangamani discussed strongly coupled $\mathcal
N=4$ Yang-Mills field theory on anti-de Sitter space, showing that
S-duality has a complicated action on the space of boundary
conditions, consistent with a dual description in terms of gravity in
a higher-dimensional AdS space. Later in the meeting, Rajesh Gopakumar
described the structure of a Hermitian matrix model in the `t Hooft
limit, showing how it could be related through Belyi maps to a
topological string theory where amplitudes are localised at specific
points in a moduli space of Riemann surfaces.

There has recently been intense interest in applications of AdS/CFT to
condensed matter systems, and there were two talks about the extension
of AdS/CFT to systems with non-relativistic scaling symmetry. Jelle
Hartong discussed correlation functions on the spacetime dual to field
theories with Schrodinger symmetry. K. Narayan showed that spacetimes
with Lifshitz symmetry can be realised in string and M theory, using a
new approach which constructs the Lifshitz spacetime by Kaluza-Klein
reduction from a solution which is a deformation of AdS. This is an
important achievement, as previous attempts to realise such an
embedding more directly failed.

Another interesting attempt to extend AdS/CFT is the Kerr-CFT
correspondence, which studies the near-horizon geometry of rotating
black holes and attempts to construct a dual CFT description. Andy
Strominger discussed  work which aims to construct an
example of such a duality within string theory, using charged black
holes in five dimensions and deforming away from a supersymmetric
solution. Alejandra Castro also talked about Kerr black holes,
describing the broken conformal symmetry hidden in the wave equation
for scalar fields on the black hole background. I gave a talk on the
similar near-horizon limit for BTZ black holes; here the CFT
description is clearer, and progress has been made on interpreting the
near-horizon limit, but there are important differences from the
Kerr/CFT case.  Glenn Barnich described another extension of AdS/CFT
ideas, considering the asymptotic isometries of asymptotically flat
spacetimes; the novelty in his approach was to allow transformations
which are not globally well-defined.

There were two talks in the meeting on the worldvolume theories of
branes in M theory. Sunil Mukhi showed that the
Bagger-Lambert-Gustavsson theory, which is supposed to be related to
M2-branes, can be related to the Yang-Mills theory on a D2-brane by
giving a vacuum expectation value to a scalar in the BLG theory. Later
in the meeting, Costas Papageorgakis described work attempting to use
three-algebras to construct worldvolume theories for multiple
M5-branes.

Returning to deep mathematical subjects, Atish Dabholkar gave a talk
on black hole entropy and mock modular forms. He explained the
properties of mock modular forms in detail - despite being first
mentioned by Ramanujan more than 50 years ago, they have only recently
been understood - and described the relation to the entropy of black
holes in theories with $\mathcal{N}=4$ supersymmetry. Bengt Nilsson
gave a talk on the use of automorphic forms to capture information on
instanton corrections in string theory.

There were two talks on supergravity in the latter part of the week:
Amitabh Virmani discussed the use of three-dimensional
solution-generating transformations to obtain new solutions in
five-dimensional supergravity, and Kelly Stelle described a new
construction of chiral supergravity theories in six and seven
dimensions.  Albion Lawrence overcame a chest infection to give a talk
on large-field inflation, arguing that in a theory where the scalar
field has a periodic identification which is only violated by the
potential, corrections to the theory can remain under control over a
range of field values much larger than the Planck scale.

There was a prize for the best presentation by a younger person at the
meeting, which was awarded to Alejandra Castro for her talk on hidden
conformal symmetry. Congratulations again to her.

There was one free afternoon to explore the surrounding area, which
offered a range of activities from boating on the lake to hiking and
cycling in the surrounding mountains. The conference banquet was held
at an outdoor restaurant a short walk away, which offered delicious
food in a more rustic setting - a change from the elegant three-course
meals we were getting all week at the conference center!

\vfill\eject

\section*{\centerline
{Condensed Matter at AdS/CFT/Strings 2010}}
\addtocontents{toc}{\protect\medskip}
\addcontentsline{toc}{subsubsection}{
\it Condensed Matter at AdS/CFT/Strings 2010, 
by Christopher Herzog}
\parskip=3pt
\begin{center}
Christopher Herzog, Princeton University 
\htmladdnormallink{cpherzog-at-princeton.edu}
{mailto:cpherzog@princeton.edu}
\end{center}

There has been gathering excitement in the string theory community that field theory/string theory dualities (often called the AdS/CFT correspondences) may be able to say something concretely useful about strongly interacting condensed matter systems, an excitement which manifested during two gatherings this past year --- ``The AdS/CFT Workshop: Nuclear and Many Body Physics'' hosted by the Princeton Center for Theoretical Science (March 11--12) and ``Strings 2010'' hosted by Texas A\&M (March 14--19).  These field theory/string theory dualities posit that certain strongly interacting field theories are actually secretly theories of classical gravity.  Thus, the computations driving progress in understanding these strongly interacting field theories may be of a certain interest to experts in general relativity.

The idea to use field theory/string theory dualities to explore condensed matter systems grew out of an earlier effort to use these same dualities to examine the quark-gluon plasma 
produced now by the Relativistic Heavy Ion Collider (RHIC)  at Brookhaven National Labs and in the future, at higher temperatures, by the Large Hadron Collider (LHC) at CERN.
The dualities provide some evidence that the viscosity of strongly interacting plasmas should have a universal and small value $\eta = \hbar s / 4 \pi k_B$ where $s$ is the entropy density. This value is consistent with experimental measurements at RHIC.  Along with some other results from AdS/CFT, the result has led several nuclear physicists to take part in the theoretical developments.  At the AdS/CFT workshop, theorists Dmitri Kharzeev, Derek Teaney, Jorge Noronha, Dam Son and experimentalists Peter Jacobs and Barbara Jacak all gave talks.
The observation about the viscosity continues to generate new work and ideas.  Rob Myers
attempted to go beyond the strong coupling limit on the gravity side of the duality by adding higher curvature corrections to the gravity action and relating causality constraints on the coefficients to bounds on $\eta$.

Now however there is growing interest in applying these dualities to understanding certain strongly interacting systems in condensed matter physics, for example superconductors and superfluids.  
At the AdS/CFT workshop, condensed matter theorist Subir Sachdev described some models of fermion systems which might be particularly well adapted to a dual gravitational formulation.  Among those with a particle theory background, Hong Liu showed how a solution to the Dirac equation in a charged black hole background with a negative cosmological constant can be used to calculate a fermionic Green's function in a dual strongly interacting field theory.  For a certain choice of mass and charge of the fermion, the Green's function is that of the ``marginal Fermi liquid'', a phenomenological form for the correlator that can produce the famed linear growth in the resistivity of the cuprate superconductors with temperature in their normal phase, a linear growth that has evaded explanation for decades.   While Liu's work requires a fine tuning of the fermion parameters,
it is nevertheless an intriguing new approach to an old and difficult problem.

Sean Hartnoll talked about going beyond the classical gravity approximation to include some gravitational loop effects.  Field theories with a classical gravity dual are typically both strongly interacting and have a large number of colors $N$.  Gravity loops correspond to $1/N$ corrections on the field theory side.  Hartnoll was interested in particular in $1/N$ corrections to the free energy which allowed him to see de Haas-van Alphen oscillations  in the field theory.  An interesting intermediate result was a new reformulation of the free energy as a sum over quasinormal modes in the black hole space-time.

While ``Strings 2010'' contained a large variety of talks on different subjects, there were a couple of sessions devoted to applying AdS/CFT to condensed matter systems.  Two opposing points of view could be distinguished.  To understand the microscopic degrees of freedom in the field theory, the gravity model has to be a low energy limit of a string theory.  Despite this limitation, many have decided to work with a gravity model whose string theory embedding is yet unknown --- a bottom-up or phenomenological approach that gives more flexibility.  Others have opted to work hard to find a stringy embedding that puts the results on a firmer footing.  An important bottom up model is the holographic superconductor described in talks by Gary Horowitz and myself.  The gravitational dual of the superconducting phase transition is an instability for a charged black hole in a space-time with negative cosmological constant to develop scalar hair.  Jerome Gauntlett described promising new top-down results that suggest ways of embedding this model in string theory.  Other excellent talks related to this program were given by Hirosi Ooguri, Shamit Kachru, Igor Klebanov, and Joe Polchinski.  A talk which perhaps deserves special mention was given by Andy Strominger.  Inspired by work of Hong Liu and others solving the Dirac equation in this AdS/CFT context, Strominger described a picture of Kerr black holes surrounded by a Fermi sea.

It remains to be seen whether the AdS/CFT approach can make a prediction for a real world condensed matter system.  Indeed, there are some difficult questions that need to be answered: What is the counter-part of large $N$ for a superconductor?  How important is it to have a string theoretic understanding of the duality?  Nevertheless, AdS/CFT is an exciting new tool for investigating a class of strongly interacting systems, and, flipping our perspective, condensed matter systems may one day have the potential to shed light on classical and quantum gravity.

\end{document}